\renewcommand{\vec}[1]{\mathbf{#1}}
\documentclass[aps,superscriptaddress,showpacs,letterpaper,twocolumn]{revtex4-1}
\usepackage[letterpaper]{geometry}
\usepackage{graphicx} 
\usepackage{amsmath}
\usepackage{color}  

\begin{document}

\title{Magic-wavelength optical traps for Rydberg atoms}

\author{S. Zhang}
\affiliation{Department of Physics, 1150 University Avenue, University of Wisconsin, Madison, Wisconsin 53706}
\author{F. Robicheaux}
 \affiliation{Department of Physics, Auburn University, Auburn, Alabama 36849-5311}
\author{M. Saffman}
\affiliation{Department of Physics, 1150 University Avenue, University of Wisconsin, Madison, Wisconsin 53706}
 \email{msaffman@wisc.edu}

\date{\today}

\begin{abstract}
We propose blue-detuned optical traps that are suitable for trapping of both ground state and Rydberg excited atoms.
Addition of a background compensation field or suitable choice of the trap geometry provides a magic trapping  condition for ground and Rydberg atoms at the trap center.
Deviations from the magic condition at finite temperature are calculated. Designs that achieve less than 200 kHz differential trap shift between Cs ground and $125s$ Rydberg states for $10 ~\mu\rm K$ Cs atoms are presented.  
Consideration of the trapping potential and photoionization rates suggest that these traps will be useful for quantum information experiments with atomic qubits. 
\end{abstract}


\maketitle

\section{Introduction}
While  ground state neutral atoms interact only weakly due to small van der Waals and magnetostatic dipolar interactions
several recent experiments have shown that Rydberg excitation can be used to turn on strong interactions suitable for quantum gates and entanglement generation
\cite{Wilk2010,Isenhower2010,Zhang2010}. Following these developments  Rydberg mediated quantum gates\cite{Lukin2001} are currently being studied intensively as a route to scalable quantum information processing\cite{Saffman2010}.

Recent quantum gate experiments have used Rb atoms that are laser cooled and then transferred into red detuned far-off resonance optical traps (FORTs). Red detuned traps are adequate for ground state atoms but they present several difficulties for experiments that rely on Rydberg excitation. The trapping light photoionizes  Rydberg atoms with typical photoionization rates for few mK deep traps that can be faster than radiative decay rates\cite{Saffman2005a,Potvliege2006,Tallant2010}, and so photoionization presents a limit to the usable Rydberg lifetime.  Furthermore the differential light shift of the Rydberg and ground states results in a position dependent Rydberg excitation energy, unless the atoms are cooled to the motional ground state of the trapping potential. Variations in the excitation energy impact the detuning of pulses used for gate operations, and thus degrade gate fidelity. 
To get around these limits the trap light is turned off during the gate sequence, and then turned on again afterwards. Provided the atoms are sufficiently cold, and the Rydberg gate lasts only a few $\mu\rm s$, turning the trap on and off does not lead to appreciable heating or atom loss out of the trap. 

In a multi-qubit experiment we envision an array of optical traps, each holding a neutral atom. In most implementations using lattices, or trap arrays generated with diffractive optics, it is not possible to control the trap intensity on a site by site basis. It would therefore be necessary to turn off all traps whenever any atom is Rydberg excited. This is problematic for implementations with many qubits and it is therefore of interest to find traps that work for both ground and Rydberg state atoms. Since the Rydberg polarizability is that of a free electron and is negative, a stable trap must be a dark region surrounded by light, and the trap wavelength should be chosen so also the ground state polarizability is negative. In the alkali atoms this implies tuning to the blue of one or both of the first resonance lines\cite{Saffman2005a}. Although a blue detuned trap can be attractive for both ground and Rydberg atoms,
trap depth matching is still an issue due to state dependent differences in the magnitude of the  polarizability and due to the different spatial extent of the Rydberg wavefunction compared to the ground state atom. For high fidelity quantum gates we expect to access Rydberg levels with principal quantum number $n>100$\cite{Saffman2010} and it is therefore necessary to consider the local trapping potential averaged over the Rydberg electron wavefunction\cite{Dutta2000} which may extend to more than $1~\mu\rm m$ away from the nucleus. 

Several authors  have considered  low frequency electromagnetic trap designs for ground and Rydberg atoms\cite{Choi2007,*Hyafil2004,*Mozley2005,*Mayle2009,*Mayle2009b}. In this paper we show that optical frequency traps can be used for both ground and Rydberg state atoms, and that position dependent differential light shifts can be minimized in what we refer to as ``quasi-magic" trap geometries.   
In Sec. \ref{sec.traps} we present three alternative designs for blue detuned optical traps. 
 In Sec. \ref{sec.ponderomotive} we calculate the Rydberg trapping potential and identify magic trapping conditions. 
Representative numbers are given for Cs atoms. Photoionization rates are calculated in Sec. \ref{sec.pi} and we conclude in Sec. \ref{sec.conclusions}.

\section{Bottle Beam Optical Traps}
\label{sec.traps}

\begin{figure}[!t]
  \centering
\includegraphics[width=0.45\textwidth]{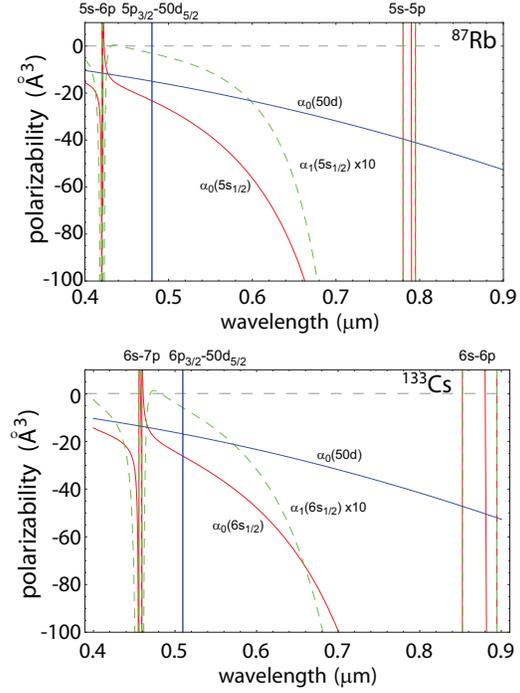}
  \caption{(color online) Scalar polarizability of ground and Rydberg states of Rb and Cs (solid lines) and 
vector polarizability of the ground state (dashed lines).}
  \label{fig.alpha}
\end{figure}

Wavelength regions where the ground and Rydberg state polarizabilities are the same sign are to the blue of the first resonance lines in alkali atoms. Calculated polarizability curves for the heavy alkalis Rb and Cs are shown in Fig. \ref{fig.alpha}. 
The curves for the $50d$ Rydberg state are within a few percent of the value found from the free electron polarizability $\alpha_e=-\frac{e^{2}}{m_{e}\omega^{2}}$, except near the resonance with the first excited $p$ level. 
We see that for both elements there is a matching wavelength to the red of the second resonance lines at approximately 430 nm for Rb and 
470 nm for Cs. The ground state vector polarizabilities are also very small at this wavelength which implies small rates for hyperfine or Zeeman state changing Raman transitions. Although these wavelengths might therefore appear attractive for trapping ground and Rydberg states they are not useful due to the need to account for the different spatial extent of the ground and Rydberg wavefunctions. As we will see in Sec. \ref{sec.ponderomotive} it is preferable to work at longer wavelengths for which the ground and Rydberg polarizabilities are both negative, but the ground state polarizability is much larger in magnitude than that of the Rydberg state. 

\begin{figure}[!t]
  \centering
  \includegraphics[width=0.45\textwidth]{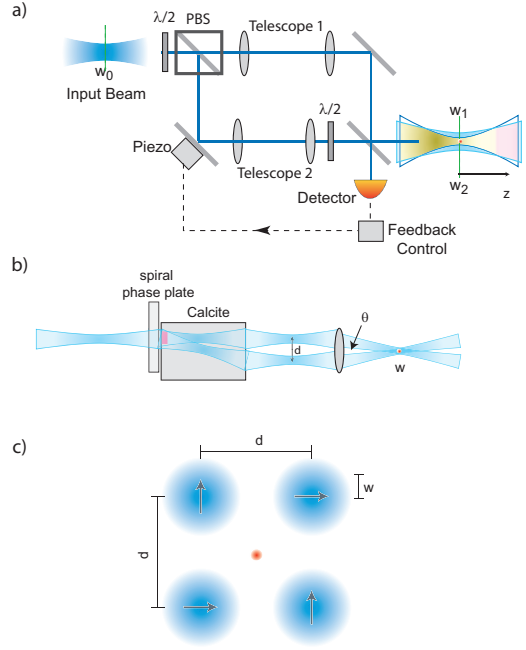}
 \caption{\label{fig:bob_setup}(color online) Setup of blue detuned dipole traps: a) Gaussian interference BoB\cite{Isenhower2009},
 b) crossed vortex BoB\cite{Fatemi2007}, and c) Gaussian lattice. }
\end{figure}

Several methods have been used to produce bottle
beam traps (BoBs) that have an intensity null surrounded
by light in all directions \cite{Kuga1997,*Ozeri1999,*Arlt2000,*Kulin2001,*Yelin2004,*Xu2010}, \cite{Fatemi2007},   \cite{Isenhower2009}. We have investigated in detail the three 
configurations shown in Fig.  \ref{fig:bob_setup}. The Gaussian interference BoB in Fig. \ref{fig:bob_setup}a)   makes use of the interference of two TEM$_{00}$ Gaussian beams with different waist sizes $w_{1}$, $w_{2}$ \cite{Isenhower2009}. The crossed vortex BoB\cite{Fatemi2007} in Fig.  \ref{fig:bob_setup}b is formed by  two Laguerre-Gaussian $L_{0}^{1}$ beams with orthogonal polarizations that cross with an angle of $2\theta$. 
We have recently demosntrated trapping of single ground state Cs atoms in both of these BoB traps\cite{SZhang2011}.

The third dipole trap \ref{fig:bob_setup}c is created by four parallel tightly focused TEM$_{00}$ Gaussian beams. The four beams with a waist size $w_{0}$ are spaced on a square with sides  $d$. Each beam has the same polarization as its diagonal neighbor and has orthogonal polarization to that of its nearest neighbors. This polarization configuration minimizes the effects of interference. Both the waist size $w_{0}$ and beam spacing $d$ are on the $\mu m$ scale. The overlap of the four beams forms a potential barrier around the center of the square in the $x-y$ plane. Diffractive spreading of the Gaussians also creates a trapping potential along $z$,  thus forming a 3D BoB trap. This latter configuration is of particular interest for forming tightly packed BoB arrays. 

For each trapping geometry atom localization near the trap center can be quantified by an expansion of the  potential
$U({\bf r})=-\frac{1}{2\epsilon_{0}c}\alpha I({\bf r})$ near the trap center. Here $\alpha$ is the scalar polarizability
and $I$ is the intensity at position $\bf r$. The intensity distributions for the different trap configurations are calculated in Appendix A. For the Gaussian interference BoB we find near the origin
\begin{subequations}
\begin{eqnarray}
    U(x,0,0)&=&-\frac{\alpha P_{1}(w_{1}^{2}-w_{2}^{2})^{2}}{\pi\epsilon_{0}c w_{1}^{6}w_{2}^{4}}x^{4}+O(x^{6})
\\
    U(0,0,z)&=&-\frac{\alpha\lambda^{2} P_{1}(w_{1}^{2}-w_{2}^{2})^{2}}{\pi^{3}\epsilon_0 c w_{1}^{6}w_{2}^{4}}z^{2}+O(z^{4})
\end{eqnarray}
\label{eq.interferenceBoB}
\end{subequations}
where $\alpha$ is the scalar polarizability of the atom, $\lambda$ is the trapping wavelength, and $P_{1}, P_{2}=(w_2/w_1)^2P_1$ are the powers of the beams with waists $w_{1}, w_{2}$ repectively. The trapping potential is axially symmetric and quartic in the $xy$ plane and quadratic along $z$.
The total trap power used in Fig. \ref{fig:bob_setup} is  $P=P_{1}+P_{2}$.

For the crossed vortex BoB an expansion about the origin yields 
\begin{subequations}
\begin{eqnarray}
    U(x,0,0)&=&-\frac{2\alpha P\cos^{2}\theta}{\pi\epsilon_{0} c w^{4}}x^{2}+O(x^{4})\\
U(0,y,0)&=&-\frac{2\alpha P}{\pi\epsilon_{0} c w^{4}}y^{2}+O(y^{4})\\
U(0,0,z)&=&-\frac{2\alpha P\sin^{2}\theta}{\pi \epsilon_{0} c 
w^{4}}z^{2}+O(z^{4})
\end{eqnarray}
\label{eq.vortexBoB}
\end{subequations}
where $P$ is the total power of the two beams, and $w$ is the focused waist size. This trap is quadratic in all directions.\\

For the Gaussian array BoB the expansion along $x$ and $z$ is 
\begin{subequations}
\begin{eqnarray}
    U(x,0,0)&=&-U_0 e^{-\frac{d^{2}}{w^{2}}}\left(1-\frac{2w^{2}-d^{2}}{w^{4}}x^{2}\right) + O(x^{4})\nonumber\\
\\
U(0,0,z)&=&-U_0 e^{-\frac{d^{2}}{w^{2}}}\left[1-\frac{\lambda^{2}(w^{2}-d^{2})}{\pi^{2}w^{6}}z^{2}\right] 
\hspace{-.1cm}+ \hspace{-.1cm} O(z^{4})\nonumber\\
\end{eqnarray}
\label{eq.gaussianBoB}
\end{subequations}
with $U_0=\frac{8\alpha P}{\pi \epsilon_{0} c w^{2}}$. Here $P$ is the power of each beam in Fig. \ref{fig:bob_setup}c. In an array implementation each beam is shared between four neighboring trapping sites, so a total power of only $P$ per site is required(this neglects a small correction due to the rows at the edge of the array).

%
\begin{figure}[!t]
\includegraphics[width=0.5\textwidth]{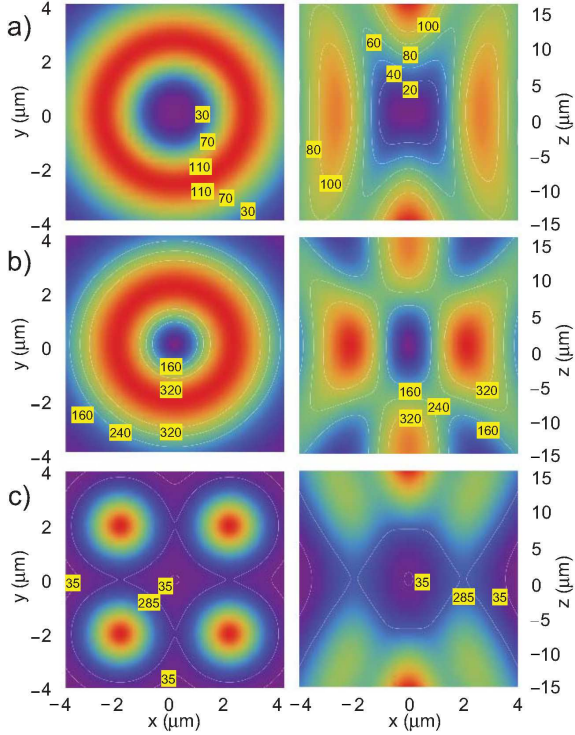}
  \caption{(color online) Trapping Depth of blue detuned dipole traps in the $x-y$ (left column)  and $x-z$ (right column) planes for Cs 6s, $\lambda=780~\rm nm$, $\alpha=-235. \times 10^{-24}~\rm cm^3$, power $P=50~\rm mW$ for  a) Gaussian interference BoB $w_{1}=2~\mu\rm m$, $w_{2}=3.78~\mu\rm  m$, b) crossed vortex BoB $w=3~\mu\rm  m$, $\theta=8.6^{o}$, and c) Gaussian lattice trap $w=1.5~\mu\rm  m$, $d=4~\mu\rm  m$.}
  \label{fig:bob_depth}
\end{figure}

Trapping potentials of the three dipole trap configurations for the ground state of Cs are plotted in Fig.~\ref{fig:bob_depth} and the trap oscillation frequencies along different axes are listed in Table \ref{tab:trap_freq}. We see that all three designs provide transverse oscillation frequencies of a few tens of kHz and longitudinal oscillation frequencies of a few kHz. The vortex and Gaussian lattice traps result in quite similar frequencies and trapping depth. The Gaussian interference BoB is about $3\times$ shallower for the same optical power and has the poorest axial confinement. 
 
\begin{table*}[!ht]
\centering
\caption{
\label{tab:trap_freq}Oscillation frequencies and the trap potential at the lowest saddle point  for the BoB traps, each with the same total power of 50 mW, with trap parameters from  Fig.\ref{fig:bob_depth}.}
\begin{tabular}{|p{0.10\textwidth}|c|c|c|c|}
\hline
 design&$\omega_{x}/2\pi$ (kHz)&$\omega_{y}/2\pi$ (kHz)&$\omega_{z}/2\pi$ (kHz)&$U/k_{B}~(\mu\rm K)$\\
\hline 
Gaussian Interference&62.5\footnotemark[1]&62.5\footnotemark[1]&0.315&60 \\
\hline 
crossed Vortex&29.4&29.8&4.42&225\\
\hline 
Gaussian Lattice &15.4&15.4&2.79&256\\
\hline
\end{tabular}
\footnotetext[1]{The Gaussian interference BoB is anharmonic in the radial direction. The vibration frequency was calculated by setting the particle energy to $1/10$ of the trapping potential.}
\end{table*}

\section{Ponderomotive Potential of Trapped Rydberg Atoms}
\label{sec.ponderomotive}

\begin{figure}[!t]
\includegraphics[width=0.5\textwidth]{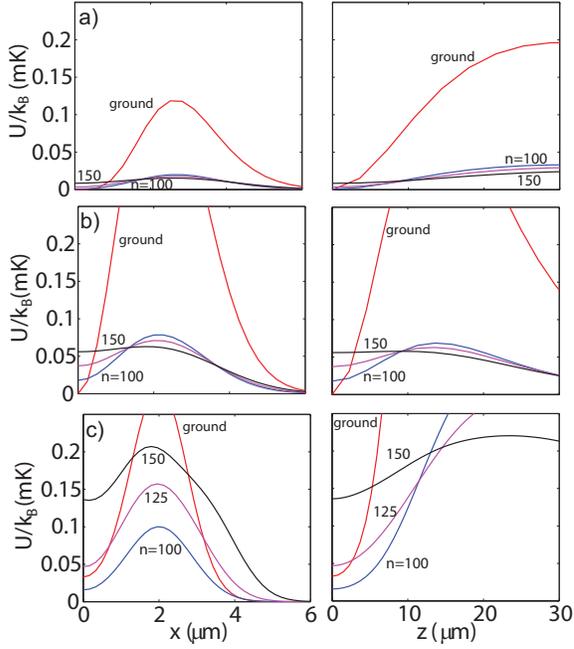}
\caption{(color online) Potential energy of Cs ground and $ns$ Rydberg states with $n=100,125,150$ in a) Gaussian interference BoB, b) vortex BoB, and c) Gaussian lattice BoB. Trap parameters the same as in  Fig.~\ref{fig:bob_depth}.}
\label{fig:Ryd_poten}
\end{figure}

High $n$ Rydberg atoms with $n>100$ have electron wavefunctions that are comparable in spatial extent to the 
trap potentials shown in Fig. \ref{fig:bob_depth}.
The AC Stark shift of Rydberg atoms can therefore no longer be approximated by $U=-\frac{1}{2\epsilon_{0}c}\alpha I$, with $I$ the local intensity at the nucleus. We need to consider the ponderomotive energy of Rydberg atoms in a field of varying intensity. 
The ponderomotive shift is the time averaged kinetic energy of a free electron in an oscillating electric field. For a field of the form $E\cos(\omega t)$, the ponderomotive energy is 
$$
    U_{P}=\frac{e^{2}|E|^{2}}{4m_{e}\omega^{2}}\nonumber
$$
where $-e$ and $m_{e}$ are the electron charge and mass respectively.
Using $ I=\frac{\epsilon_{0}c}{2}|E|^{2}$
where $c$ is  the speed of light we can write  the ponderomotive energy of a free electron  as 
$$
    U_{P}=\frac{e^{2}}{2\epsilon_{0}c m_e\omega^{2}}I.
$$
Then the Hamiltonian of a Rydberg atom in an oscillating electromagnetic field can be written as 
$$
    {H_{F}+U_{P}(\vec{R}+\vec{r})}\psi(\vec{r};\vec{R})=E_{R}(\vec{R})\psi(\vec{r};\vec{R}),
$$
where $\vec{R}$ is the center of mass coordinate of the atom, and $\vec{r}$ is the coordinate of the electron relative to the center of mass. 
Using first order perturbation theory, and supposing there is no degeneracy involved, the energy shift of a Rydberg atom in  state $j$ is \cite{Dutta2000}
\begin{eqnarray}
    \Delta E_{Rj}(\vec{R})&=&\int d^{3}rU_{P}(\vec{R}+\vec{r})|\psi^{0}_{j}(\vec{r};\vec{R})|^{2}\nonumber\\
&=&\frac{e^{2}}{2\epsilon_{0} c m_e\omega^{2}}\hspace{-.1cm}\int d^{3}r I(\vec{R}+\vec{r})|\psi^{0}_{j}(\vec{r};\vec{R})|^{2}.
\label{eq.UP}
\end{eqnarray}
This expression is valid provided the ponderomotive potential varies over distance scales that are larger than the wavelength of the Rydberg electron. This is well satisfied for the potentials we consider. At $n=150$ the electron wavelength is about 50 nm which is less than 10 \% of the
wavelength of the light creating the trap. 
In addition it is necessary that the ponderomotive shift is everywhere small compared to the energy spacing of Rydberg levels.  For the $150s$ state, which is the highest we consider in the examples below, the closest state is 
$146f_{7/2}$ which is 1.6 mK away. Looking at Fig. \ref{fig:Ryd_poten} the largest ponderomotive energy seen by a $150s$ atom for the traps  we are considering is about $200~\mu\rm K$. The ratio of energy scales would thus imply a higher order correction
$\sim  0.12$.   

In fact this naive estimate is overly pessimistic since the coupling between $ns$ and $(n-4)f$ is strongly suppressed by the trap geometry. 
The dipole 
traps in Fig. \ref{fig:bob_setup} all have  spatial reflection symmetry so that the coupling matrix 
elements between the $ns$ and $(n-4)f$ Rydberg states are exactly zero when the 
atom is at the origin. For the low atom temperatures expected for Cs, 
the atom will be very near the center of the trap and
thus this coupling will be strongly suppressed. The next closest states 
are in the (n-4) degenerate manifold ($l = 4, 5, ...$) which are separated 
from the $150s$ state by about $5~\rm mK$ which is more than a factor of 25 
larger than the light shift. This will lead to a second order 
perturbative correction to the energy shift of approximately 1 part in 
25. 
Since the Rydberg level spacing scales as $1/n^3$, with $n$ the principal quantum number, the error will be even smaller for lower levels.

We will calculate the wavefunctions $\psi_j^0$ using a  model pseudo potential method. The potential form adopted here is \cite{Robicheaux1997}
\begin{equation}
V_{l}(r)=-\frac{Z_{l}(r)}{r}-\frac{\alpha_{d}}{2r^{4}}
\left[1-e^{-(r/r_{c})^3}\right]^{2}+\frac{l(l+1)}{2r^{2}}\label{eq.Vl}
\end{equation}
where $Z_{l}(r)=1+36e^{-\alpha_{l}^{(1)}r}+\alpha_{l}^{(2)}re^{\alpha_{l}^{(3)}r}$. $\alpha_{d}=15.81$, $r_{c}=2.0$, and all the other parameters are listed in Table \ref{tab:potential_parms}.
\begin{table}[!ht]
\centering
\caption{Parameters for the Cs model potential (\ref{eq.Vl}).}
\begin{tabular}{|c|c|c|c|c|c|}
\hline
 $l$&0&1&2&3&4+\\
\hline 
$\alpha_{l}^{(1)}$&3.49625&3.73801&3.45092&3.43592&3.43592\\
\hline
$\alpha_{l}^{(2)}$&9.57499&9.56664&9.52285&9.54285&9.54285\\
\hline
$\alpha_{l}^{(3)}$&1.41409&1.34016&1.58147&1.62147&1.62147\\
\hline
\end{tabular}
\label{tab:potential_parms}
\end{table}
To verify our calculation of the wavefunctions, we reproduced the planewave photoionization cross sections listed in \cite{Saffman2005a}. The fine structure of Cs adds  less than 0.1\% correction to the ponderomotive energy shifts, so we  ignore  fine structure corrections in this paper.\\

Figure \ref{fig:Ryd_poten} gives sample  calculation results for $ns$ Rydberg levels with $n=100,125,150$. 
We  see that as $n$ increases the effective trapping potential gets smaller and smaller. This is because the large electron wavefunction averages over the intensity distribution of the trap according to Eq. (\ref{eq.UP}) which washes out the potential minimum.  If the trap parameters are not chosen correctly, as is the case in Fig. \ref{fig:Ryd_poten}b), the trap could be repulsive for high $n$  even though $\alpha_e$ is negative. Even when the trap is attractive for Rydberg states the ground to Rydberg trap  shift for an atom at ${\bf R}=0$ is not negligible. This shift  increases 
 with $n$ and is  proportional to the light intensity.
In an experiment with an array of traps this would imply that the Rydberg excitation energy would vary from site to site due to intensity variations across the array. To minimize this effect we seek trap parameters for which the ${\bf R}=0$ trap induced shift vanishes. We will refer to this in what follows as ``quasi-magic" trapping. A quasi-magic trap will give an intensity independent excitation shift for atoms 
at the trap center (or for atoms in the motional ground state with slightly different compensation parameters) and only a small shift for sufficiently cold atoms. We quantify the notion of small in the following section.

\section{Magic Condition for Zero Temperature Atoms}
\label{sec.magic}

Inspection of Fig. \ref{fig.alpha} shows that apart from wavelengths that are very close to the second resonance lines 
 the magnitude of the ground state polarizability is larger than that of the Rydberg state. Conversely Fig. \ref{fig:Ryd_poten} shows that the trapping potential at ${\bf R}=0$ is larger for Rydberg states than for ground states. This implies that we can balance the $\bf R=0$ trap shifts by adding a constant background intensity  that will shift the ground state potentials more than the Rydberg state potentials. With the correct background intensity $I_{\rm m}$ the differential shift will vanish. This is the quasi-magic trapping condition. Note that if we were to use the wavelengths in Fig. \ref{fig.alpha} where the ground and Rydberg polarizabilities are equal we would have to add a relatively large background intensity. At $\lambda=780~\rm nm$ the ground state polarizability $\alpha$  is about  $5.4\times$ larger than that of the Rydberg state $\alpha_e$ which reduces the power requirement for the background beam by this factor. It is possible to work even closer to the first resonance line where $\alpha/\alpha_e$ is even larger, but decoherence rates associated with photon scattering and differential hyperfine shifts\cite{Saffman2005a,Kuhr2005}
increase correspondingly. We have therefore chosen 
$780~\rm nm$ for Cs as a viable working wavelength.

\begin{figure}[!t]
\centering
\includegraphics[width=0.48\textwidth]{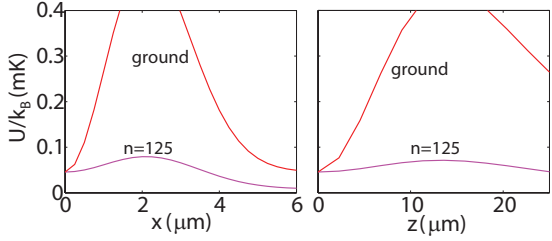}
\caption{(color online) Energy shift compensation for the crossed vortex BoB with a planewave of intensity $I_{m}=128~\mu\rm W/\mu m^{2}$, $w=3~\mu\rm m$, $\theta=8.6^{\rm o}$, and $P=50~\rm mW$.}
\label{fig:magic}
\end{figure}

Using the  ground state light shift 
\begin{equation}
    \Delta U_{g}=-\frac{\alpha_{g}}{2\epsilon_{0}c}[I_{\rm BoB}(\vec{R})+I_{\rm m}(\vec{R})],\nonumber
\end{equation}
and the Rydberg state shift
\begin{equation}
\begin{split}
    \Delta U_{R}&=\frac{e^{2}}{2\epsilon_{0}c m_e\omega^{2}}\nonumber\\
    &\int d^{3}r [I_{\rm BoB}(\vec{R}+\vec{r})+I_{\rm m}(\vec{R}+\vec{r})]|\psi^{0}_{j}(\vec{r};\vec{R})|^{2}\nonumber
    \end{split}
\end{equation}
the quasi-magic condition is simply $\Delta U_g=\Delta U_R.$
Figure \ref{fig:magic} shows an example of such a magic condition for the crossed vortex BoB. 

\begin{figure}[!t]
\centering
\includegraphics[width=0.5\textwidth]{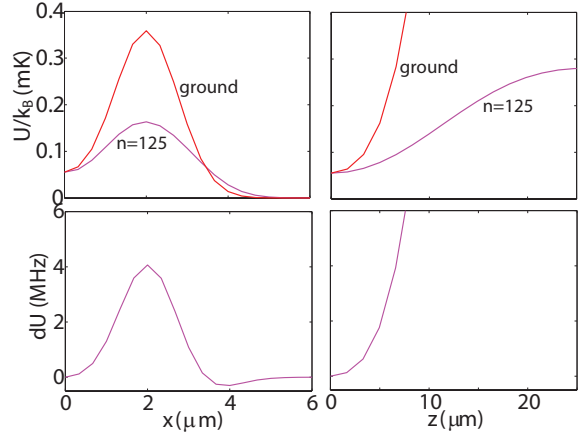}
\caption{(color online) Trapping potential (top row) and shift difference (bottom row) between Cs $6s$ and  $125s$ for a self-magic Gaussian lattice trap with $\lambda=780~\rm nm$, $d=4~\mu\rm  m$, $w=1.57~\mu\rm  m$, and $P=50~\rm mW$.}
\label{fig:self_magic}
\end{figure}

Although the additional power required for matching is small
for a single site, the additional light requirement becomes substantial if we consider a $100\times 100$ or $1000\times 1000~\mu\rm m^2$ array. The Gaussian lattice design presents an interesting alternative since the light intensity is naturally not zero at the trap center. 
 The ${\bf R}=0$ intensity changes as we vary the waist size or separation of the beams, and by judicious choice of parameters we can achieve the matching condition without adding any additional plane wave. Note that the compensating intensity is in this case not uniform but is spatially varying. 
 Figure \ref{fig:self_magic} shows such a self magic condition for $n=125$.

\begin{figure}[htbp]
\centering
\includegraphics[width=0.4\textwidth]{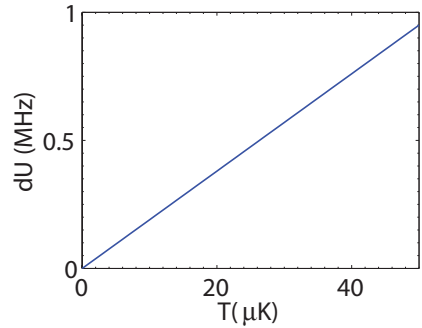}
\caption{(color online) Average transition shift between ground and 125s states of Cs in Gaussian lattice trap with $d=4~\mu\rm  m$, $w=1.57~\mu\rm  m$, $P=50~\rm mW$, and $U_{\rm trap}=k_B\times 300~\mu\rm K$.}
\label{fig:dU_vs_T}
\end{figure}

For a ground state atom with a low temperature, we can estimate the average trap induced shift between ground and Rydberg states by $<dU>=\frac{1}{2}\sum_{i=x,y,z}{dU_{ii}(0,0,0)\left< r_{i}^{2}\right>}$, where  the mean square position of the atom found from 
the Virial theorem is $<r_{i}^2>=\frac{k_{B}T}{2\partial_{ii}U_{g}}$, $dU_{ii}=\partial_{ii}(U_R-U_g)$ and  $U_g, U_R$
are the ground and Rydberg state trapping potentials. Figure \ref{fig:dU_vs_T} shows that the transition shift decreases nearly linearly with decreasing atom temperature. This shift would be below 0.2 MHz for an atom temperature of 10 $\mu \rm K$ which is readily achieved using polarization gradient cooling of Cs.

\section{Photoionization Rate}
\label{sec.pi}

In this section we calculate the photoionization rate of Rydberg atoms in a BoB trap. Since the Rydberg electron is not exposed to a uniform intensity field the photoionization calculation must be modified accordingly. 
The local photoionization rate $R$ scales as 
$    R=\sigma \frac{I}{\hbar \omega}$ with $\omega$ the photon frequency.
The cross section is \cite{Gallagher1994}
\begin{equation}
    \sigma=2\pi^2 \frac{\hbar^{2}}{m_{e}^{2}c^{2}\alpha}\frac{df_{if}}{dE}, \nonumber
\end{equation}
 $\alpha$ is the fine structure constant and 
the derivative of the oscillator strength with respect to the energy is
\begin{equation}
 \frac{d f_{if}}{dE}=\frac{2\hbar}{m\omega}\frac{1}{2l_{i}+1}\sum_{m_i}\sum_{l_{f},m_{f}}|\langle f|\vec{A}\cdot\vec{p}|i\rangle|^{2}.
\label{eq:fif1}
\end{equation}
where the initial state $|i\rangle=|n_{i},l_{i},m_{i}\rangle$ is a Rydberg state with principle quantum number $n_{i}$, and the  final state  $|f\rangle=|E_{f},l_{f},m_{f}\rangle$ is a continuum state with energy $E_{f}=E_{\omega}+E_{R}$, with $E_\omega=\hbar\omega$ the photon energy. The magnitude of $\bf A$ is normalized to  unit peak intensity. 
Even though 
$kr\sim kn^2 a_0\gg 1$ is large for our parameters  we may ignore high powers of $r$ in the expansion of $\bf A$ when  calculating the matrix element since, even though the electron's wavefunction is comparable in size to 
the photon's wavelength, the photon absorption takes place near the 
nucleus\cite{Fano1985}.  
The quadrupole term is included due to the small electric dipole transition rate for $s$ state atoms near the center of a dark trap. 
We tested the matrix 
element calculation using the full multipole operator and verified that 
only the dipole and quadrupole terms gave a substantial contribution.

For a planewave field polarized in the $x$ direction $\vec{A}=e^{i\vec{k}\cdot\vec{r}}\hat{x}$, and Eq. (\ref{eq:fif1}) can be approximated by
\begin{equation}
\begin{split}
\frac{df_{if}}{dE}\approx &\frac{2\hbar}{m\omega}\frac{1}{2l_{i}+1}\\
\times&\sum_{m_i}\sum_{l_{f},m_{f}}|\langle f|p_{x}+ik_{x}xp_{x}+ik_{y}yp_{x}+ik_{z}zp_{x}|i\rangle|^{2}.
\end{split}
\label{eq:fif2}
\end{equation}
Using the following relations 
\begin{eqnarray} 
xp_{x}&=\frac{im}{2\hbar}(Hxx-xxH)+\frac{1}{2}i\hbar\nonumber\\
yp_{x}&=\frac{im}{2\hbar}(Hxy-xyH)-\frac{1}{2}l_{z}\nonumber\\
zp_{x}&=\frac{im}{2\hbar}(Hxz-xzH)+\frac{1}{2}l_{y}.\nonumber
\end{eqnarray}
and dropping the magnetic dipole terms, which give no contribution to the photoionization rate, Eq. (\ref{eq:fif2}) becomes
\begin{eqnarray}
\frac{df_{if}}{dE}&\approx&\frac{2m\omega}{\hbar}\frac{1}{2l_{i}+1}\nonumber\\
&\times & \sum_{m_i}\sum_{l_{f},m_{f}}\left|\langle f|x+\frac{ik_{x}}{2}x^{2}+\frac{ik_{y}}{2}xy+\frac{ik_{z}}{2}xz|i\rangle \right|^{2}.\nonumber
\end{eqnarray}

For a spatially inhomogeneous   field like the Gaussian lattice BoB we decompose into planewaves as 
\begin{eqnarray}
    A(\vec{r})&=&\frac{1}{(2\pi)^{3}}\int d^{3}\vec{k}\, g_{\vec{k}}e^{i\vec{k}\cdot\vec{r}},\nonumber\\
    g_{\vec{k}}&=&\int d^{3}\vec{r}\, A(\vec{r})e^{-i\vec{k}\cdot\vec{r}}\nonumber.
\end{eqnarray}
The oscillator strength derivative can then be written as
\begin{equation}
\begin{split}
\frac{d    f_{if}}{dE}\approx&\frac{2m\omega}{(2\pi)^6\hbar}\frac{1}{2l_{i}+1}\sum_{m_i}\sum_{l_{f},m_{f}}\\
&\left|\int d^{3}\vec{k}\, g_{\vec{k}}\left(\langle x\rangle
    +\frac{ik_{x}}{2}\langle xx\rangle+ \frac{ik_{y}}{2}\langle xy\rangle+\frac{ik_{z}}{2}\langle xz\rangle\right)\right|^{2}\nonumber.
\end{split}
\end{equation}
To evaluate the matrix elements the wavefunctions are calculated with the same method as in Section \ref{sec.ponderomotive}.
The radial part of the continuum state is normalized to 
$$
\phi_{E_{f},l_{f}}\rightarrow\sqrt{\frac{2m}{\pi \hbar^{2}k_{e}}}\sin(k_{e}r+\delta), \textrm{as } r\rightarrow\infty.
$$ 
where $k_{e}$ is the wavenumber of the free electron and $\delta$ is the continuum state phase shift.

\begin{figure}[!t]
\includegraphics[width=0.5\textwidth]{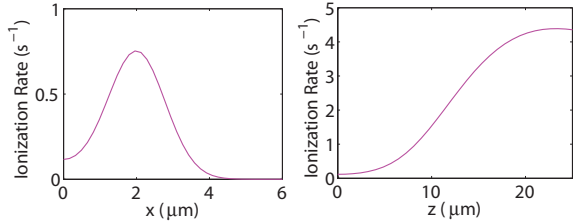}
\caption{(color online) Photoionization rate for 125s Cs in a $780~\rm  nm$ self-magic Gaussian lattice dipole trap, $w=1.57~\mu \rm m$, $d=4 ~\mu \rm m$, $P=50~\rm mW$, and $U_{\rm trap}=k_B\times 300~\mu\rm K$.
}
\label{fig:ionRate}
\end{figure}

Figure \ref{fig:ionRate} shows the photoionization rate for a $125s$ state Cs atom in a Gaussian lattice trap which satisfies the self magic condition of Fig. \ref{fig:self_magic}. The quadrupole term gives less than 3\% correction to the final result shown in the plot due to the non-zero light intensity at the trap center.
The photoionization rate is also substantially smaller than the room temperature radiative decay rate of the same Rydberg state which is about $1800~\rm s^{-1}.$ 
Nevertheless the  room temperature
photoionization rate of the $125s$ state is dominated by 
 blackbody radiation which gives a rate of about\cite{Beterov2009c} $20~\rm s^{-1}$ at
 $300~\rm K$. Only at cryogenic temperatures less than $10~\rm K$ does the trap light induced rate at trap center seen in Fig. \ref{fig:ionRate} dominate over the blackbody rate.  

\section{Discussion and Conclusions}
\label{sec.conclusions}

In summary we have presented three designs for   blue detuned dipole traps that are capable of trapping both ground and Rydberg state atoms.  Using visible or near-infrared trapping wavelengths, and alkali atoms with temperatures $<100~\mu\rm K$, these traps are capable of $\mu\rm m$ scale atomic localization in three dimensions.
We have calculated the ponderomotive potential energy of trapped Rydberg atoms, the importance of which has been demonstrated  in recent experiments \cite{Younge2010}, and shown that it is possible to match the ground and Rydberg state trap shifts
for atoms at the center of the trap.

One attractive feature of these optical traps is that they can be replicated easily in two dimensions with a diffractive beam splitter. In this way the traps could be used in experiments that require control over individual sites of a closely spaced two-dimensional atomic array as in\cite{Weitenberg2011}. This type of holographically replicated 
and projected array has the interesting feature compared to more traditional optical lattices that the
position of each trapping site does not depend on a relative phase between two interfering beams. This suggests the potential for better long term stability compared to optical lattice implementations. 

A particular feature of the Gaussian array BoB trap (Fig. \ref{fig:bob_setup}c) is that a periodic array of Gaussian beams creates an array of dark traps with the same periodicity, without any extra confining walls. This approach would enable quasi-magic trapping of $125s$ atoms on a lattice with $4~\mu\rm m$ periodicity as detailed in Sec. \ref{sec.magic}.
Conversely the Gaussian interference and crossed vortex BoB traps (Fig. \ref{fig:bob_setup}a,b) have a confining wall around each trap site so that there would be two confining walls between each site in a replicated array. This implies an approximately 50\% larger trap period for quasi-magic trapping of $125s$ atoms, which would reduce the number of sites per unit area by more than a factor of two.

As we have shown in Sec.. \ref{sec.magic} quasi-magic ground-Rydberg trap shift  matching can be achieved either by adding a uniform background field to the trap designs that have zero intensity at the trap center (Fig. \ref{fig:bob_setup}a,b), or by careful choice of the trap parameters of the Gaussian lattice trap 
(Fig. \ref{fig:bob_setup}c) which has a finite intensity at the trap center.
 These quasi-magic traps have no intensity dependent shift 
for atoms at the trap center (or in the motional ground state), but do show shifts at finite  temperature. We have shown in Sec. \ref{sec.magic} that the finite temperature shifts can be limited to $\sim 200~\rm kHz$ for $10~\mu\rm K$ Cs atoms, and would be even less for colder atoms.
Trap shift matching is important for high fidelity Rydberg mediated quantum gates\cite{Saffman2010}, furthermore 
this method may also be relevant for high accuracy control of black body radiation shifts\cite{Beloy2006} in optical transition atomic clocks.
 In principle it may be possible to improve upon our results by designing a compensating field with the correct spatial shape such that not only the differential shift at trap center, but also higher spatial derivatives of the differential shift are canceled. We leave this as a challenge for future work.

\section{Acknowledgments}
SZ and MS received support from the IARPA MQCO program through ARO contract W911NF-10-1-0347, DARPA,
and NSF awards PHY-1005550, PHY-0969883.  
FR was supported by the NSF under grant no 0969530.



%

\appendix
\section{Calculation of trap intensity distributions}

In this appendix we document the calculation steps used to derive the intensity distributions leading to the  trapping potentials of Eqs. (\ref
{eq.interferenceBoB},\ref{eq.vortexBoB},\ref{eq.gaussianBoB}) for the Gaussian beam interference BoB, crossed vortex BoB, and Gaussian array BoB respectively. 

\subsection{Gaussian Interference BoB}

The two Gaussian beams have a phase difference of $\pi$ after the Mach-Zehnder interferometer, and the on-axis intensities are set equal by putting 
 $P_1/w_1^2=P_2/w_2^2$ with $P_{1,2}$ the power and $w_{1,2}$ the beam waist. 
The combined intensity of the BoB trap is
\begin{widetext}
\begin{equation}
\begin{split}
I(\rho,z)=\frac{2P_1}{\pi w_1^2}\left|\frac{w_1}{w_1(z)}e^{-\frac{\rho^{2}}{w_{1}^{2}(z)}}e^{\imath[kz+k\frac{\rho^2}{2R_{1}(z)}-\eta_{1}(z)]}
-\frac{w_2}{w_2(z)}e^{- \frac{\rho^{2}}{w_{2}^{2}(z)}}e^{\imath[kz+k\frac{\rho^2}{2R_{2}(z)}-\eta_{2}(z)]}\right|^2
\nonumber
\end{split}
\end{equation}
\end{widetext}
where from the properties of TEM$_{00}$ Gaussian beams $z_{R1,2}=\pi w_{1,2}^2/\lambda$, $w_{1,2}(z)=w_{1,2}\sqrt{1+(\frac{z}{z_{R1,2}})^2}$, $R_{1,2}(z)=z+\frac{z_{R1,2}^2}{z}$, $\eta_{1,2}(z)=\arctan(\frac{z}{z_{R1,2}})$ and $\rho^2=x^2+y^2$. Multiplying by the polarizability to convert to energy units and Taylor expanding about the origin gives Eqs. (\ref{eq.interferenceBoB}).

\subsection{Crossed Vortex BoB}

The intensity of a Laguerre-Gaussian beam can be written as
\begin{widetext}
\begin{equation}
I_{l,p}(\rho,z)=I_{0}\left(\frac{C_{lp}}{w(z)}\right)^{2}\left(\frac{2r^{2}}{w^{2}(z)}\right)^{|l|}e^{-\frac{2\rho^{2}}{w^{2}(z)}}
\left[L_{p}^{|l|}\left(\frac{2\rho^{2}}{w^{2}(z)}\right)\right]^{2}\nonumber
\end{equation}
\end{widetext}
where $I_{0}=\frac{P}{w_{0}^{2}}$, $C_{lp}=\sqrt{\frac{2p!}{\pi (l+p)!}}$, 
$w(z)=w_0\sqrt{1+(\frac{z}{z_{R}})^{2}}$, and $z_{R}=\frac{\pi w_{0}^{2}}{\lambda}$. For the crossed vortex BoB we are using, $l=1$, $p=0$. 

The BoB trap is created by passing two orthogonally polarized beams with separation $d$ through a lens of focal length $f$. After the focusing lens, the two beams are rotated by $\pm\theta=\pm\arctan(\frac{d}{2f})$ in the $x,z$ plane. For not too large angles such that we can neglect local polarization changes due to the beam focusing  the BoB  intensity is given by 

\begin{equation}
I(x,y,z)=I_{1,0}(\rho_+,z_+)+I_{1,0}(\rho_-,z_-)\nonumber
\end{equation}
with $\rho_\pm=\sqrt{y^{2}+(x\cos\theta\pm z\sin\theta)^{2}}$ and $z_{\pm}=z\cos\theta \mp x \sin\theta$.
 Multiplying by the polarizability and Taylor expanding about the origin gives Eqs. (\ref{eq.vortexBoB}).

\subsection{Gaussian Lattice}

Each unit cell of the Gaussian beam lattice has the same polarization on the upper right and lower left corners and an opposite polarization on the lower right and upper left corners as shown in Fig. \ref{fig:bob_setup}c). We therefore add the fields from the beams centered at opposite corners, and then add the two intensities. This can be written as  
\begin{widetext}
\begin{eqnarray}
I(x,y,z)=\frac{2P}{\pi w_0^2}\left[\left|E(x-d/2,y-d/2,z)+E(x+d/2,y+d/2,z)\right|^2\right. \nonumber \\
+\left. \left|E(x+d/2,y-d/2,z)+E(x-d/2,y+d/2,z)\right|^2\right] \nonumber
\end{eqnarray}
\end{widetext}
where 
each side of the unit cell has length $d$, $P$ is the power of each beam,  and the unity normalized field distribution of each beam is 
\begin{equation}
E(x,y,z)=\frac{w_{0}}{w(z)}e^{-\frac{x^{2}+y^2}{w^{2}(z)}}e^{\imath[ kz+k\frac{x^{2}+y^{2}}{2R(z)}-\eta(z)]}. \nonumber
\end{equation}
Taylor expanding the potential about the center of the unit cell at $x=y=0$ gives Eqs. ({\ref{eq.gaussianBoB}).

\end{document}